\documentclass[twocolumn,showpacs,preprintnumbers,floatfix]{revtex4}
\usepackage{graphicx}
\usepackage{dcolumn}
\usepackage{bm}

\begin{document}

\title{Narrowband frequency tunable light source of continuous quadrature entanglement}
\author{Christian Schori}
\author{Jens L. S\o rensen}
\email{jls@ifa.au.dk}
\author{Eugene S. Polzik}
\affiliation{QUANTOP - Danish National Research Foundation Centre for Quantum Optics,\\
Institute of Physics and Astronomy, University of Aarhus, DK-8000, Denmark.}
\date{\today }

\begin{abstract}
We report the observation of non-classical quantum correlations of
continuous light variables from a novel type of source. It is a frequency
non-degenerate optical parametric oscillator below threshold, where signal
and idler fields are separated by 740MHz corresponding to two free spectrum
ranges of the parametric oscillator cavity. The degree of entanglement
observed, - 3.8 dB, is the highest to-date for a narrowband tunable source
suitable for atomic quantum memory and other applications in atomic physics.
Finally we use the latter to visualize the Einstein-Podolsky-Rosen paradox.
\end{abstract}

\maketitle

\section{Introduction}

Quantum interface between light and atomic systems requires narrowband
frequency tunable sources of light possessing various non-classical
features. Examples of sources which are in demand include narrowband
tunable single photon sources and narrowband tunable sources of
Einstein-Podolsky-Rosen entangled light. Both types of sources can be
constructed using a non-degenerate optical parametric oscillator below
threshold (ND-OPO). However, the vast majority of current experiments with
entangled light is carried out with single pass downconversion sources with
their inherent low spectral brightness. The first EPR-type experiment using a
narrowband light source \cite{ou92a,ou92b} was also the first demonstration
of entanglement for continuous variables. Recently another experiment
demonstrating the EPR state for solitons has been carried out \cite{silberhorn01}
although in this experiment the bandwidth of the correlated fields was
very wide. The first demonstration of a frequency tunable source of a
non-classical narrowband light source possessing entanglement between the
modes at 852 nm and 917 nm was reported in \cite{georgiades95}. A
potentially tunable source of entangled beams at the same wavelength but in
different spatial modes was used in the continuous variables
teleportation experiment \cite{furusawa98}. In the latter experiment a
degenerate OPO pumped in two opposite directions was utilized to
generate two squeezed beams which were mixed on a beamsplitter to produce
two entangled beams with the degree of entanglement around -1 dB. In the
present paper we demonstrate the entanglement of a new type. Namely two
symmetric longitudinal modes of the OPO cavity which are shown to possess the
quadrature phase entanglement as suggested in the original paper by M. Reid 
\cite{reid89} where entanglement of the OPO output has been discussed for
the first time.

\section{ND-OPO theory}

In this section we present a model for calculating quantum fluctuations of
the ND-OPO output modes. Our treatment is based on \cite{reid89} with
experimentally relevant generalizations: the two modes do not
necessarily experience the same losses in the resonator and the output
coupling rates can be different too. The ND-OPO is described as a moderate
finesse optical resonator containing a $\chi ^{(2)}$ nonlinear medium. This
resonator is assumed to be resonant with two field modes $\widetilde{a}_{+}$
and $\widetilde{a}_{-}$, henceforth denoted the signal and idler modes
respectively. In terms of frequency the signal and idler modes are assumed
to be an integer number of free spectrum ranges higher and lower relative to
the degenerate frequency set by the classical external pump field $\widetilde{\alpha }$.\ That is, energy conservation requires that the sum of
the frequencies of the resonating fields must equal the pump field frequency 
$\omega _{+}+\omega _{-}=\omega _{p}$, so that $\omega _{\pm }=\omega
_{p}/2\pm \nu \omega _{FSR}$, where $\omega _{FSR}$ is the frequency of a
free spectrum range and $\nu $ is an integer. The mode operators obey the
usual bosonic commutation relation 
\begin{equation}
\left[ \widetilde{a}_{n}(t),\widetilde{a}_{m}^{\dagger }(t)\right]
=\delta _{nm},\;\;\left\{ n,m\right\} =\left\{
+,-\right\}  \label{modecomm}
\end{equation}

\subsection{Dynamics}

The starting point for our treatment of the ND-OPO is the Hamiltonian 
\begin{equation}
H=H_{free}+H_{int}+H_{res}
\label{htot}
\end{equation}
where $free$, $int$ and $res$ describes the free, noninteracting OPO\ cavity
fields, the nonlinear interaction between the fields and the coupling of the
fields to the external reservoir of vacuum modes respectively. The first two
Hamiltonians take the form 
\begin{equation}
H_{free}=\hbar \omega _{+}\widetilde{a}_{+}^{\dagger }\widetilde{a}
_{+}+\hbar \omega _{-}\widetilde{a}_{-}^{\dagger }\widetilde{a}_{-}
\label{hfree}
\end{equation}
and 
\begin{equation}
H_{int}=i\hbar g \left(\widetilde{\alpha }e^{i\chi }\widetilde{a}
_{+}^{\dagger }\widetilde{a}_{-}^{\dagger }-\widetilde{\alpha }^{*}e^{-i\chi }
\widetilde{a}_{+}\widetilde{a}_{-}\right)
\label{hint}
\end{equation}
where $g$ is the coupling strength, $\widetilde{\alpha}=\alpha \exp{(-2i\omega t)}$ is the classical pump field and $\chi$ is the pump field phase.
From (\ref{htot}), (\ref{hfree}) and (\ref{hint}) we can derive the Heisenberg-Langevin equations of motion for the
cavity fields 
\begin{equation}
\begin{array}{r}
\frac{d}{dt}\widetilde{a}_{\pm }=-i\omega _{\pm }\widetilde{a}_{\pm }+g
\widetilde{\alpha }e^{i\chi }\widetilde{a}_{\mp }^{\dagger }-\gamma _{\pm }
\widetilde{a}_{\pm } \\ 
+\sqrt{2\kappa _{\pm }}\widetilde{a}_{\pm }^{in}+\sqrt{2(\gamma _{\pm
}-\kappa _{\pm })}\widetilde{b}_{\pm }^{in}
\end{array}
\label{eqmot}
\end{equation}
where the $\widetilde{a}^{in}$ fields describe the vacuum fluctuations leaking into the
resonator through the main coupler and the $\widetilde{b}^{in}$ fields enter the ND-OPO via
residual leakage. $\kappa _{\pm }$ define the leak rate through the main
coupler at frequencies $\omega _{\pm }$ and $\gamma _{\pm }$ define the
overall leak rates. Our observables of interest are the slowly varying envelopes of the field quadratures
\begin{equation}
q_{\pm}(\theta)=\widetilde{a}_{\pm} e^{i(\omega_{\pm}t-\theta)}+\widetilde{a}_{\pm}^{\dagger} e^{-i(\omega_{\pm}t-\theta)}
\label{quaddef}
\end{equation}
The pump field we parameterize in terms of its threshold value $\alpha=\varepsilon \alpha_{th}$, where $\alpha_{th}=\sqrt{\gamma_+ \gamma_-}/g$. This results in the equations of motion for the subharmonic quadrature operators
\begin{equation}
\begin{array}{r}
\frac{d}{dt}q_{\pm }(\theta )=\varepsilon \sqrt{\gamma _{+}\gamma _{-}}
q_{\mp }(\chi -\theta )-\gamma _{\pm }q_{\pm }(\theta ) \\ 
+\sqrt{2\kappa _{\pm }}q_{\pm }^{in}(\theta )+\sqrt{2(\gamma _{\pm }-\kappa
_{\pm })}p_{\pm }^{in}(\theta )
\label{quadmotion}
\end{array}
\end{equation}
Here we have defined the slowly varying vacuum field quadratures $q_{\pm }^{in}(\theta
)=\widetilde{a}_{\pm }^{in}e^{i(\omega_{\pm}t-\theta) }+(\widetilde{a}_{\pm }^{\dagger })^{in}e^{-i(\omega_{\pm}t-\theta)}$ and $
p_{\pm }^{in}(\theta )=\widetilde{b}_{\pm }^{in}e^{i(\omega_{\pm}t-\theta) }+(\widetilde{b}_{\pm }^{\dagger
})^{in}e^{-i(\omega_{\pm}t-\theta) }$. The phase dependence of $q^{in}$ and $p^{in}$ will be
left out in the following since no phase can be assigned to the vacuum state
and hence the properties of the '$in$' quadrature phase operators is independent of
the choice of $\theta$.

From (\ref{quadmotion}) we see that only the quadratures with phases $\phi
_{\pm }=\chi /2\pm \theta $ will interact. Hence we find that the ND-OPO
intracavity fields are described by a continuum of pairwise coupled
equations of motion of the form

\begin{equation}
\begin{array}{r}
\frac{d}{dt}q_{\pm }(\phi _{\pm })=\varepsilon \sqrt{\gamma _{+}\gamma _{-}}
q_{\mp }(\phi _{\mp })-\gamma _{\pm }q_{\pm }(\phi _{\pm }) \\ 
+\sqrt{2\kappa _{\pm }}q_{\pm }^{in}+\sqrt{2(\gamma _{\pm }-\kappa _{\pm })}
p_{\pm }^{in}
\end{array}
\label{couplequad}
\end{equation}

\begin{figure}
\includegraphics{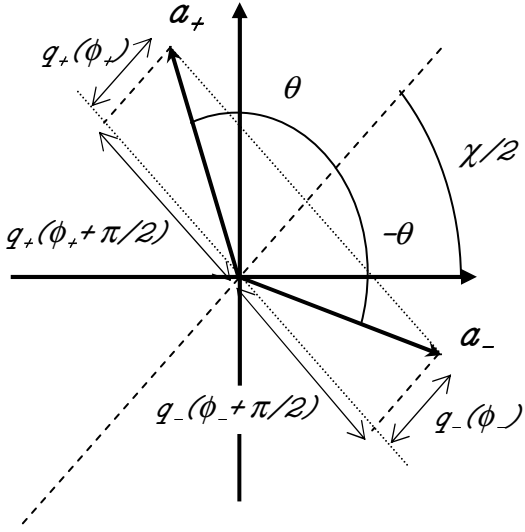}
\caption{\label{fig:phasor} The parametric downconversion seen in a phasor diagram. In a frame rotating at the degeneracy frequency $a_+$ is rotating clockwise and $a_-$ is rotating counterclockwise with relative phases such that the two phasors always add up to be parallel to the dashed line.}
\end{figure}

Before moving on to solve (\ref{couplequad}) we make a few observations
about the quadrature phase operators. Using (\ref{modecomm}) we find that
these operators obey the following commutation relations 
\begin{equation}
\left[ q_{n}(\phi ),q_{m}(\phi ^{\prime })\right] =2i\sin (\phi ^{\prime
}-\phi )\delta _{nm},\left\{ m,n\right\} =\left\{ +,-\right\}
\label{quadcomm}
\end{equation}
meaning that the combined signal/idler quadratures describing cross
correlations of the ND-OPO output commute as 
\begin{widetext}
\begin{equation}
\left[ q_{+}(\phi _{+})+q_{-}(\phi _{-}),q_{+}(\phi _{+}+\psi )+q_{-}(\phi
_{-}+\psi )\right] =  
\left[ q_{+}(\phi _{+})-q_{-}(\phi _{-}),q_{+}(\phi _{+}+\psi )-q_{-}(\phi
_{-}+\psi )\right] =4i\sin \psi
\label{croscomm}
\end{equation}
\end{widetext}
Hence the combinations $\{q_{+}(\phi _{+})+q_{-}(\phi _{-}),q_{+}(\phi
_{+}+\pi /2)+q_{-}(\phi _{-}+\pi /2)\}$ and $\{q_{+}(\phi _{+})-q_{-}(\phi
_{-}),q_{+}(\phi _{+}+\pi /2)-q_{-}(\phi _{-}+\pi /2)\}$ form conjugate
observables. As we will see below the fluctuations of the quadrature
difference $q_{+}(\phi _{+})-q_{-}(\phi _{-})$ can become very small in the
ND-OPO output as compared to a two mode coherent state output. This means
that $q_{+}(\phi _{+}+\pi /2)+q_{-}(\phi _{-}+\pi /2)=q_{+}(\phi
_{+}^{\prime })-q_{-}(\phi _{-}^{\prime })$, with $\phi _{\pm }^{\prime
}=\chi /2\pm (\theta +\pi /2)$, also can become a quiet observable since $\phi _{\pm }$ is defined for any arbitrary value of $\theta $. From the
commutation relation (\ref{croscomm}) we then conclude that $q_{+}(\phi
_{+})+q_{-}(\phi _{-})$ and $q_{+}(\phi _{+}+\pi /2)-q_{-}(\phi _{-}+\pi /2)$
must become correspondingly noisy observables. This is illustrated in the
phasor diagram of Fig. \ref{fig:phasor}. 
In the parametric down conversion the phase is conserved, meaning that the
signal and idler fields are created symmetrically around the axis making an
angle of $\chi /2$ with the abscissa in phase space. This axis illustrates the pump phase. In the rotating frame oscillating at $\omega _{p}/2$ the
signal field will rotate clockwise and the idler will rotate
counterclockwise in such a way that their sum always will be parallel to the pump
axis. Hence we find that the difference of projections of signal/idler
fields onto the pump axis will tend to cancel on average. This difference is
described by the cross correlation of quadratures $q_{+}(\phi
_{+})-q_{-}(\phi _{-})$. Also the \textit{sum} of projections onto the axis
perpendicular to the pump axis will cancel. This observable is described by $q_{+}(\phi _{+}+\pi /2)+q_{-}(\phi _{-}+\pi /2)$.

\subsection{Quantum fluctuations}

The solutions of (\ref{couplequad}) are found by Fourier transforming and
solving the resulting algebraic equations 
\begin{equation}
\mathcal{D}\vec{q}=\mathcal{A}\vec{q}^{in}+\mathcal{B}\vec{p}^{in}
\label{algebraeq}
\end{equation}
where we have defined 
\begin{equation}
\begin{array}{c}
\mathcal{D}=\left( 
\begin{array}{cc}
\gamma _{+}-i\Omega & -\varepsilon \sqrt{\gamma _{+}\gamma _{-}} \\ 
-\varepsilon \sqrt{\gamma _{+}\gamma _{-}} & \gamma _{-}-i\Omega
\end{array}
\right) ,\,\mathcal{A}=\left( 
\begin{array}{cc}
\sqrt{2\kappa _{+}} & 0 \\ 
0 & \sqrt{2\kappa _{-}}
\end{array}
\right) , \\ 
\mathcal{B}=\left( 
\begin{array}{cc}
\sqrt{2(\gamma _{+}-\kappa _{+})} & 0 \\ 
0 & \sqrt{2(\gamma _{-}-\kappa _{-})}
\end{array}
\right)
\end{array}
\label{matdef}
\end{equation}
and 
\begin{equation}
\vec{q}=\left( 
\begin{array}{c}
q_{+}(\phi _{+}) \\ 
q_{-}(\phi _{-})
\end{array}
\right) ,\,\vec{q}^{in}=\left( 
\begin{array}{c}
q_{+}^{in} \\ 
q_{-}^{in}
\end{array}
\right) ,\,\vec{p}^{in}=\left( 
\begin{array}{c}
p_{+}^{in} \\ 
p_{-}^{in}
\end{array}
\right)
\end{equation}
now are the Fourier transformed quadrature phase operators. From the
solution of (\ref{algebraeq}) we can find the fields escaping the ND-OPO\
through the main coupler via the boundary condition on the coupler 
\begin{equation}
\vec{q}^{out}=\mathcal{A}\vec{q}-\mathcal{I}\vec{q}^{in}  \label{boundary}
\end{equation}
These are given by 
\begin{equation}
\vec{q}^{out}=\left[ \mathcal{AD}^{-1}\mathcal{A-I}\right] \vec{q}^{in}+
\mathcal{AD}^{-1}\mathcal{B}\vec{p}^{in}  \label{outfields}
\end{equation}
where $\mathcal{I}$ is the 2$\times $2 unit matrix.\ To describe the
nonclassical correlations between the $out$ fields in (\ref{outfields}), we
define the general linear combinations of the signal and idler quadratures 
\begin{equation}
\vec{Q}=\left( 
\begin{array}{c}
Q_{+} \\ 
Q_{-}
\end{array}
\right) =\mathcal{U}\vec{q},\;\;\mathcal{U}=\left( 
\begin{array}{cc}
\cos \psi & \sin \psi \\ 
-\sin \psi & \cos \psi
\end{array}
\right)  \label{crossqdef}
\end{equation}
Hence we find the outgoing linear combinations as 
\begin{equation}
\vec{Q}^{out}=\mathcal{U}\left[ \mathcal{AD}^{-1}\mathcal{A-I}\right] 
\mathcal{U}^{T}\vec{Q}^{in}+\mathcal{UAD}^{-1}\mathcal{BU}^{T}\vec{P}^{in}
\label{generalout}
\end{equation}
where $\vec{Q}^{in}=\mathcal{U}\vec{q}^{in}$ and $\vec{P}^{in}=\mathcal{U}
\vec{p}^{in}$. For the vacuum input fields we define the standard quantum
limit (SQL) to be: 
\begin{equation}
\left\langle Q_{\pm }^{in}(-\Omega )Q_{\pm }^{in}(\Omega )\right\rangle
=\left\langle P_{\pm }^{in}(-\Omega )P_{\pm }^{in}(\Omega )\right\rangle
=\lambda (\psi )  \label{sql}
\end{equation}
and all other correlations functions of the $in$ fields are vanishing for a
vacuum state input. Clearly $\lambda $ must depend on our choice of $\psi $
since for $\psi =0$ $Q_{\pm }^{out}$ describes the fluctuations of the
individual signal and idler fields of the ND-OPO, whereas for $\psi \neq 0$ $Q_{\pm }^{out}$ describes inter-beam correlations. As we shall see below in
the case of symmetrical signal/idler losses and perfect inter-beam
correlations we measure the strongest inter-beam correlations when choosing $\psi =\pi /4$. Hence we define $\lambda (\psi =0)=1$ and $\lambda (\psi \neq
0)=2$, since in the latter case we must compare our inter beam-correlations
to a SQL given by the correlations between two independent coherent state
fields. It is worth noting here that only very close to the ND-OPO threshold
are the inter-beam quantum correlations perfect and the noise of the signal
and idler can cancel each other exactly. However the single beam noise grows
rapidly as the ND-OPO pump parameter $\varepsilon $ grows. Hence due to
imperfect quantum correlations between signal and idler fields it turns out
that since this growing noise cannot be canceled, it is more advantageous to
measure not the symmetric combination of the quadratures described by $\psi
=\pi /4$, but rather a somewhat scaled combination with $\psi =\pi /4+\delta
(\varepsilon )$\cite{reid89}.

The correlations of the outgoing quadratures $\left\langle (Q_{\pm}^{out})^2\right\rangle_{\psi}=\left\langle Q_{\pm }^{out}(-\Omega )Q_{\pm }^{out}(\Omega )\right\rangle_{\psi}$ are found from (\ref{generalout}) and (\ref{sql})
to be 
\begin{widetext}
\begin{equation}
\left( 
\begin{array}{c}
\left\langle (Q_{+}^{out})^2 \right\rangle _{\psi } \\ 
\left\langle (Q_{-}^{out})^2 \right\rangle _{\psi }
\end{array}
\right) =\lambda \left\{ 1+4\varepsilon \eta \frac{2\varepsilon \left[
\sigma \left( 
\begin{array}{c}
\cos ^{2}\psi \\ 
\sin ^{2}\psi
\end{array}
\right) +\sigma ^{-1}\left( 
\begin{array}{c}
\sin ^{2}\psi \\ 
\cos ^{2}\psi
\end{array}
\right) \right] + \left( \Delta ^{2}+1+\varepsilon ^{2}\right) \sin 2\psi
\left( 
\begin{array}{c}
1 \\ 
-1
\end{array}
\right) }{\left[ \Delta ^{2}+\left( E+\Lambda \right) ^{2}\right] \left[
\Delta ^{2}+\left( E-\Lambda \right) ^{2}\right] }\right\}
\label{genfluc}
\end{equation}
\end{widetext}
Here we have defined the normalized fluctuation frequency $\Delta =\Omega /\sqrt{\gamma _{+}\gamma _{-}}$, the signal/idler escape
efficiencies $\eta _{\pm }=\kappa _{\pm }/\gamma _{\pm }$, the generalized
OPO\ escape efficiency for the signal and idler fields $\eta =\sqrt{\eta
_{+}\eta _{-}}$, the escape efficiency asymmetry $\sigma =\sqrt{\eta
_{+}/\eta _{-}}$, the loss asymmetry function $\Lambda =(\rho +\rho ^{-1})/2$, with $\rho =\sqrt{\gamma _{+}/\gamma _{-}}$ and the effective pump
parameter $E=\sqrt{\varepsilon ^{2}+\Lambda ^{2}-1}$. From (\ref{genfluc})
we find for $\psi =0$ the single beam noise spectra 
\begin{equation}
\begin{array}{l}
\left\langle (q_{\pm }^{out})^2\right\rangle=1+\eta _{\pm }\frac{
8\varepsilon ^{2}}{\left[ \Delta ^{2}+(E+\Lambda )^{2}\right] \left[ \Delta
^{2}+(E-\Lambda )^{2}\right] }
\end{array}
\end{equation}
These are independent on $\theta$, meaning that the individual signal and
idler fields contain phase insensitive excess noise relative to the standard
quantum limit here being 1. As the threshold is approached $E\rightarrow
\Lambda$, and we find the noise to diverge at zero detuning. Naturally the
linearized approach taken here breaks down in this limit, where we must take
pump depletion into account as the ND-OPO gain becomes large and much
energy is transferred from the pump to the fluctuations of the signal/idler
fields.

By setting $\psi =\pi /4$ we find from (\ref{genfluc}) the inter-beam
correlations 
\begin{equation}
\begin{array}{l}
\left\langle (Q_{\pm}^{out})^2\right\rangle_{\pi /4} = 
 2\left\{ 1\pm 4\varepsilon \eta \frac{\Delta ^{2}+1+\varepsilon ^{2}\pm
(\sigma +\sigma ^{-1})\varepsilon }{\left[ \Delta ^{2}+(E+\Lambda )^{2}
\right] \left[ \Delta ^{2}+(E-\Lambda )^{2}\right] }\right\} 
\label{simcross}
\end{array}
\end{equation}
It is easy to check that in the case of symmetric losses $\rho =\sigma =1$, $\Lambda =1$ and $E=\varepsilon $, we recover from (\ref{simcross}) the well
known result for the cross correlations \cite{collett87} 
\begin{equation}
\left\langle (Q_{\pm}^{out})^2\right\rangle_{\pi /4}=2\left\{ 1\pm \eta \frac{4\varepsilon }{\Delta ^{2}+(\varepsilon \mp
1)^{2}}\right\} 
\label{symmopo}
\end{equation}
From (\ref{symmopo}) we find perfect inter-beam correlations at low
frequencies as the threshold is approached $\varepsilon \rightarrow 1$,
which reflects that $\psi =\pi /4$ corresponds to the natural signal/idler
linear combination to measure in the case of symmetric losses for the two
modes. 
\begin{figure}
\includegraphics{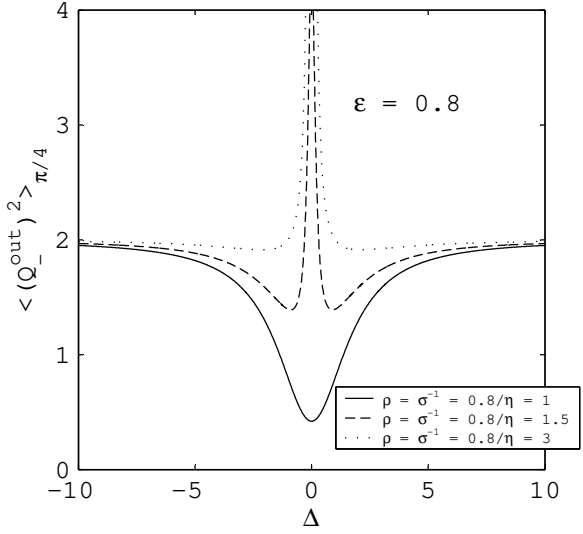}
\caption{Correlation spectra for $\kappa_+=\kappa_-$ and increasing loss asymmetry.}
\label{fig:aloss_0}
\end{figure}

When allowing for non-equal losses of signal and idler modes, we see from
Fig. \ref{fig:aloss_0} that a low frequency noise spike appears in the
spectrum of fluctuations when $\gamma _{+}$ is increased relatively to $\gamma _{-}$. With $\rho =3$ we find that the inter-beam correlations are
virtually lost. This is because the simple sums and differences of signal
idler fields, corresponding to choosing $\psi =\pi /4$, no longer are the
best combinations to measure in order to observe strong inter-beam
correlations. As we shall see by optimizing $\psi $ we can recover at least
some of the inter beam correlations around a given frequency.
\begin{figure}
\includegraphics{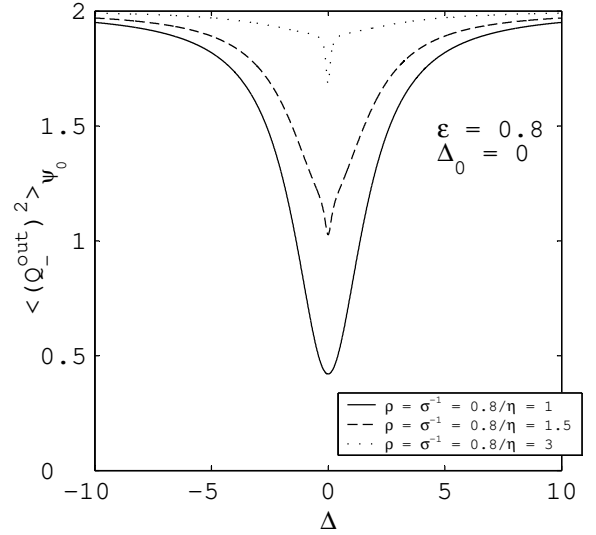}
\caption{\label{fig:aloss_1} Correlation spectra with increasing $\gamma_+$ for the optimized quadrature $Q_-^{out}$. Here $\psi_0$ has been used.}
\end{figure}

A remark should be made about losses experienced by the signal and idler
fields during propagation from the ND-OPO to the detectors. Any such losses
including non-perfect detectors, limited local oscillator overlaps etc. for
the signal/idler fields will reduce the quantum correlations on equal
footing with non-unity escape efficiencies. Hence we define overall
propagation efficiencies for the two field modes describing the survival
probability for photons in these modes from birth in the $\chi ^{(2)}$
process until detection. We denote these efficiencies $\xi _{\pm }$ for the
two modes and for later reference we define them here as 
\begin{equation}
\xi _{\pm }=\eta _{\pm }\,\eta _{\pm }^{prop}\,\eta _{\pm }^{LO}\,\eta _{\pm
}^{QE}  \label{efficiencies}
\end{equation}
Here we have defined the signal/idler propagation efficiencies $\eta _{\pm
}^{prop}$, the overlap with the local oscillators of the heterodyne
quadrature detections $\eta _{\pm }^{LO}$ and the quantum efficiency of the
detectors $\eta _{\pm }^{QE}$. In the realistic case of lossy propagation
and non-perfect detection we should replace the signal/idler escape
efficiencies in (\ref{genfluc}) with the overall efficiencies defined in (\ref{efficiencies}).

As mentioned above we can optimize the inter-beam correlations at a given
set of parameters by choosing to measure the optimized linear combination
described by $\psi =\psi _{0}$, where $\psi _{0}$ is found from 
\begin{equation}
\tan 2\psi _{0}=\frac{\Delta _{0}^{2}+1+\varepsilon ^{2}}{\left( \sigma
-\sigma ^{-1}\right) \varepsilon }  \label{optphase}
\end{equation}
Here $\Delta _{0}$ is the frequency at which we wish to minimize
fluctuations. From (\ref{optphase}) it is again clear that in the symmetric
limit where $\sigma =1$ there is no advantage of deviating from measuring
equally weighted $q_{+}^{out}(\phi _{+})$ and $q_{-}^{out}(\phi _{-})$
observables corresponding to $\psi _{0}=\pi /4$. But in general we find a
large variation of $\psi _{0}$ with $\sigma $ for small values of $\Delta
_{0}$, hence there may be a big advantage of optimizing $\psi $ when
signal/idler losses become non-symmetric. As $\Delta _{0}$ is increased the
variation of $\psi _{0}$ gets smaller since at higher frequencies the impact
of measuring unevenly weighted signal/idler amplitudes gets negligible.

\begin{figure}
\includegraphics{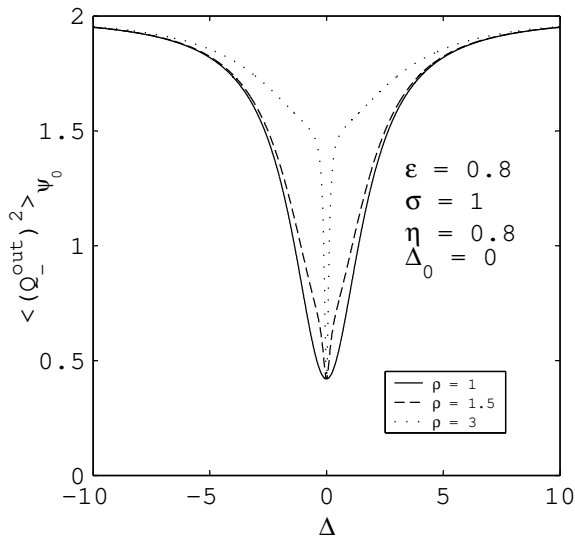}
\caption{\label{fig:aloss_2} Correlation spectra now with fixed $\sigma$ and $\eta$ and increasing $\gamma_+$. This corresponds to the situation where $\kappa_+$ grows with the same rate as $\gamma_+$.}
\end{figure}

By using $\psi =\psi _{0}$ in (\ref{genfluc}) and again varying $\gamma _{+}$ relative to $\gamma_{-}$, we
find the inter beam correlation spectra shown in Fig. \ref{fig:aloss_1}. 
These have been optimized at $\Delta _{0}=0$. Here we see that the
correlations at low frequencies persist despite the growing asymmetry of the
losses, however since the escape efficiency $\eta _{+}$ decreases with
increasing $\gamma _{+}$, we find the overall degree of inter-beam
correlations to diminish when $\rho $ deviates from 1. The importance of
measuring scaled signal and idler fields is seen more clearly from Fig. \ref{fig:aloss_2}, which corresponds to the somewhat unrealistic situation when $\rho $ is increased, but $\sigma $ and $\eta $ are left unchanged. Still
keeping $\psi $ optimized we find no loss of inter beam correlations in an
increasingly small frequency window around $\Delta =0$. Optimizing $\psi $
around a nonzero frequency is not possible without a reduction of
correlations and in general the gain in correlations by optimizing $\psi $
is marginal when $\Delta _{0}>1$. 

As a conclusion to this section we remark that not only the quantum fields
will show the behavior depicted in Fig. \ref{fig:phasor}, but also classical
fields. This means that if we inject a classical coherent field with
frequency $\omega _{+}$ and phase $\phi _{+}$ into our ND-OPO, the nonlinear
interaction will result in generation of an idler field of frequency $\omega
_{-}$ and phase $\phi _{-}$ such that $\phi _{+}+\phi _{-}=\chi $, where $\chi $ is the pump field phase. This was utilized in the experiment to gain
information about the pump phase of the ND-OPO.

\section{Experimental details}

We will now consider the experiment carried out to characterize the frequency ND-OPO described in the previous section. This was divided into three parts: First we simply measured the squeezing in the degenerate mode of the OPO. Then we investigated the signal/idler cross correlations of the ND-OPO output and finally we recorded the signal/idler phases to confirm the relationships derived above and to visualize the EPR paradox for our kind of quantum mechanically entangled light. But before considering the results we will just outline here some of the underlying experimental setup.

This was made up of a single ended optical ring resonator with a 10mm long Potassium Niobate crystal located in the smallest waist. Potassium Niobate is famous for its high $\chi^{(2)}$ nonlinearity which can be up to 8 pm/volt. However it is equally notorious due to its nonlinear losses of which Blue Light Induced InfraRed Absorption (BLIIRA) is the most harmful \cite{mabuchi94,shiv95}. In this experiment the nonlinear crystal had a single pass nonlinearity of $0.015 \pm 0.002$ W$^{-1}$ measured by second harmonic generation. The ND-OPO output coupler had a power transmission of 0.122 around the signal/idler wavelength of 860 nm. Residual losses in the ND-OPO were $0.007\pm 0.001$ excluding the BLIIRA. The cavity length was 81 cm resulting in a free spectrum range of 370 MHz. With a typical value of BLIIRA of 0.015 at 160 mW blue pump power we find the ND-OPO cavity bandwidth of 8.5 MHz FWHM.

\begin{figure}
\includegraphics{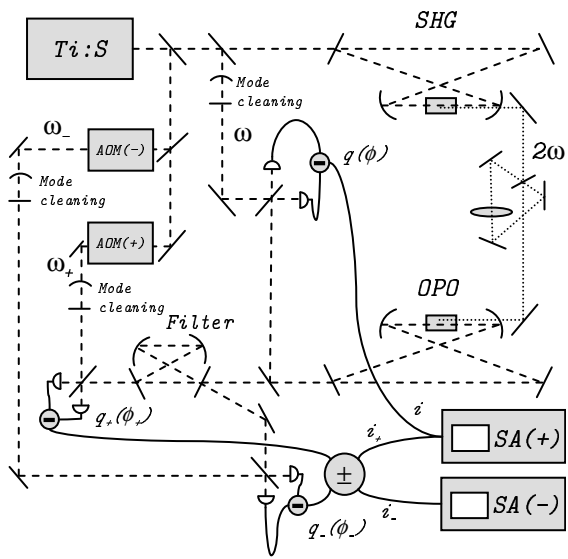}
\caption{\label{fig:setup} The experimental setup. Long dashed lines are infrared beams ($\omega_{\pm}$ and $\omega$), short dashed lines are blue beams ($2\omega$) and full lines are electrical cables. Frequency shifted local oscillators for the $\omega_{\pm}$ heterodyne detectors were generated with the acousto optical modulators AOM($\pm$). For the role of the miscellaneous cavities see text. The final photocurrents were Fourier transformed and recorded in the RF spectrum analyzers SA($\pm$).}
\end{figure}

The ND-OPO pump was produced by performing external cavity second harmonic generation (SHG) of a Titanium-Sapphire (Ti:S) laser operating at 860nm. This was done in a single ended ring resonator also with a 10 mm Potassium Niobate crystal placed in the waist. The cavity was 70 cm long and the input coupler transmitted 0.078 at 860 nm. Residual internal losses were at the level of 0.009 again excluding BLIIRA. The cavity waist was 20 $\mu$m, close to the optimum for nonlinear conversion of 16 $\mu$m \cite{boyd68}. However since the resonator was pumped with up to 800 mW of light from the Ti:S a slightly larger waist was chosen to compensate thermal lensing in the nonlinear medium. From the SHG we got up to 300 mW of blue light (430 nm) for pumping the OPO. The mode matching of the SHG Gaussian mode to the ND-OPO mode was typically 0.95 measured by the reference resonator shown in the setup on Fig. \ref{fig:setup}.

The signal and idler fields emerging from the ND-OPO to be analyzed were separated in oscillation frequency by two free spectrum ranges, that is 740 MHz. In order to separate the fields spatially we sent them onto a two ended ring resonator with a length of 20 cm and coupler transmissions of 0.078 at 860 nm. The length of this filtering cavity was stabilized to maintain resonance with $\omega_+$, such that this frequency component was transmitted. The cavity linewidth was 38 MHz and the free spectrum range 1.5 GHz, meaning that less than $10^{-3}$ of the idler field was transmitted when the filtering cavity was resonant with $\omega_+$. Mode matching and internal losses of the filtering cavity limited the transmission of the signal field to 0.95. This cavity as well as the SHG and OPO cavities were locked using the Pound-Drever-Hall FM technique \cite{bjorklund83}. The mode cleaning cavities were locked to give a constant transmission at roughly 80\% of their maximum transmission.

From the filtering cavity the signal and idler fields propagated in different directions each to be heterodyned with a frequency degenerate strong local oscillator and detected on a pair of detectors. The local oscillators were produced by sending part of the Ti:S output in double pass through a pair of acousto optical modulators driven by a 185 MHz RF source. A small portion of the up-shifted light ($\omega_+$) was injected into the filtering cavity counter propagating to the ND-OPO idler field in order to stabilize the cavity length. Additional small portions of the $\omega_{\pm}$ local oscillators were split off to have the possibility to inject them into the ND-OPO through a high reflector. The remainder of the local oscillators were passed through their respective mode cleaning cavities to ensure a clean TEM$_{00}$ mode and finally interfered with the ND-OPO outputs on separate 50/50 beam splitters. The outputs from the beam splitters were detected on four Silicon pin diodes and the photocurrents from each pair of detectors we subtracted to make a balanced measurement of the quadrature phases of the signal/idler fields. The two differential photocurrents were led into a hybrid junction producing the sum and difference of these photocurrents to be Fourier transformed in a pair of RF spectrum analyzers. The quantum efficiencies of the photodiodes were measured to be higher than 0.98 and the fringe visibilities on the 50/50 beam splitters were 0.97 and 0.98 for the $\omega_-$ and $\omega_+$ fields respectively.

As mentioned earlier we also investigated the degree of squeezing in the degenerate OPO output mode. For this purpose we inserted a mirror between the OPO and the filtering cavity. This mirror deflected the OPO output to a third heterodyne detector with a local oscillator extracted directly from the Ti:S laser. Hence the OPO output field was demodulated at exactly the frequency of degeneracy of the OPO. The fringe visibility in the degenerate heterodyne detector was 0.965 and the detector quantum efficiencies were also here higher than 0.98. The phase dependent quantum fluctuations were recorded by feeding the differential photocurrent of the detector pair directly into a RF spectrum analyzer.

\section{Results}

In order to learn about the ability of our OPO to produce non classical quantum correlations we first analyzed the degenerate output mode. Here either increased or reduced quantum fluctuations should be observed depending on the relative phase difference between the heterodyne local oscillator and the OPO output. The phase of fluctuations in the latter was determined by the pump phase. Since the same mechanisms are responsible for squeezing as well as signal/idler cross correlations the squeezing results yielded valuable information about what could be expected when observing the nondegenerate OPO modes. Next employing the filtering cavity we analyzed the two nondegenerate modes closest to the degenerate mode. Here the cross correlations between signal and idler fields were measured while we tracked the local oscillator phases. This enabled us to visualize the EPR paradox now with frequency nondegenerate quantum correlated fields as opposed to the polarization nondegenerate fields of Ou \textit{et al} \cite{ou92a}. We will discuss this in more detail below. 

\subsection{Classical gain and squeezing}

When observing squeezing we pumped the OPO with 160 mW of 430 nm light from the SHG stage. At this power we observed a BLIIRA level of $0.015\pm 0.002$ and by injecting a weak coherent field into the OPO via a high reflector we observed a maximum phase sensitive gain of 6.8. It can be shown \cite{sorensen98} that this gain is given by $(1-\varepsilon)^{-2}$, which gives us $\varepsilon=0.62$. Taking into account the blue mode matching of 0.95 this gives us an OPO threshold power of 395 mW. From $\alpha_{th}$ putting $\gamma_+=\gamma_-$ we find the power needed to drive the OPO above threshold to be
\begin{equation}
P_{th}=\frac{(\mathcal{T+L}+\mathcal{L}_P)^2}{4E_{NL}}
\label{thpower}
\end{equation}
where $\mathcal{T}$ is the OPO coupler transmission, $\mathcal{L}$ are residual OPO losses, $\mathcal{L}_P$ is BLIIRA and $E_{NL}$ is the single pass nonlinearity of the $\chi^{(2)}$ medium. From our derived threshold power we now find $E_{NL}=0.013\pm 0.001$ W$^{-1}$ in reasonable agreement with the value of $0.015\pm 0.002$ W$^{-1}$ obtained from second harmonic generation in the OPO.

\begin{figure}
\includegraphics{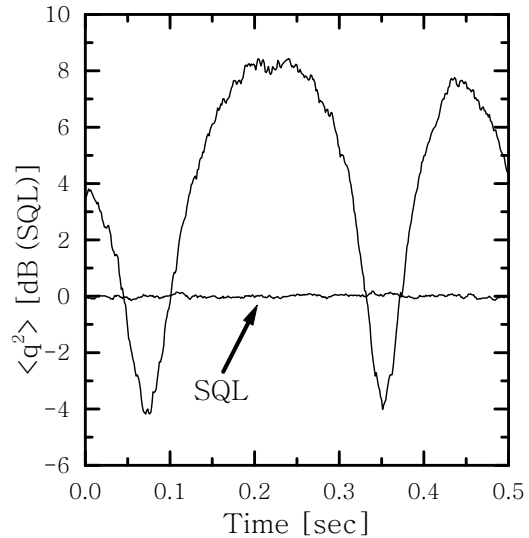}
\caption{\label{fig:sqzing} The quantum fluctuations of the degenerate OPO mode. The trace was taken with the local oscillator phase scanned. A minimum of -4.0 dB squeezing relative to the SQL was observed.}
\end{figure}

Given the measured losses of the OPO, the quantum efficiencies of the detectors, the fringe visibility in the heterodyne detection scheme and measured propagation losses of 2\%, we find using (\ref{efficiencies}) an overall detection efficiency of quantum correlations of $\xi=0.76\pm 0.02$. The squeezing balance is summarized in Table \ref{tab:sqz}.
\begin{table}
\begin{ruledtabular}
\begin{tabular}{lc}
Loss source & $\eta$ \\ 
\hline
OPO escape efficiency, $\eta$ & 0.847 \\ 
Heterodyning fringe visibility, $\eta^{LO}$ & $(0.965)^2$ \\ 
Detector quantum efficiency, $\eta^{QE}$ & 0.98 \\
Propagation efficiency, $\eta^{prop}$ & 0.98 \\
\hline
Overall detection efficiency, $\xi$ & $0.76\pm 0.02$ \\
\end{tabular}
\end{ruledtabular}
\caption{\label{tab:sqz} The limitations to the degree of squeezing and their contributions.}
\end{table}
The photocurrent resulting from the balanced heterodyne measurement was sent into a RF spectrum analyzer to be Fourier analyzed in a 30 kHz frequency window centered at 1.15 MHz. Here the quantum noise of the OPO output $\left\langle q^2 \right\rangle$ was recorded while the local oscillator phase was scanned at a rate of approximately 10 rad/sec. To establish the standard quantum limit the OPO pump was blocked and the resulting vacuum input to the heterodyne detector was analyzed to give the vacuum noise level. By blocking all inputs to the heterodyne detector the  electronic noise floor was established. This was found to be 11.76 dB below the SQL when the local oscillator power was 5 mW. A typical spectrum analyzer trace of the quantum noise is shown in Fig \ref{fig:sqzing}.
From analyzing 10 such traces we found quantum fluctuations in the squeezed quadrature of the OPO output to be at $-4.0\pm 0.1$ dB relative to the SQL and in the anti squeezed quadrature $8.0\pm 0.2$ dB of excess noise was found. Since our detectors had limited sensitivity the squeezing results were influenced by the electronic noise. After subtraction of electronic noise we found the corrected squeezing of $-4.5\pm 0.1$ dB and the anti squeezing of $8.3\pm 0.2$ dB relative to the SQL. The degree of squeezing should be compared to the theoretical value found from \cite{collett85}
\begin{equation}
\left\langle q^2 \right\rangle=1-\xi\frac{4\varepsilon}{\Delta^2+(1+\varepsilon)^2}
\label{sqzeq}
\end{equation}
Inserting our parameters in (\ref{sqzeq}) we find an expected degree of squeezing of $-5.2\pm 0.3$ dB which we believe is in reasonable agreement with the observed value.

\subsection{Entangled state of two modes}

\begin{figure}
\includegraphics{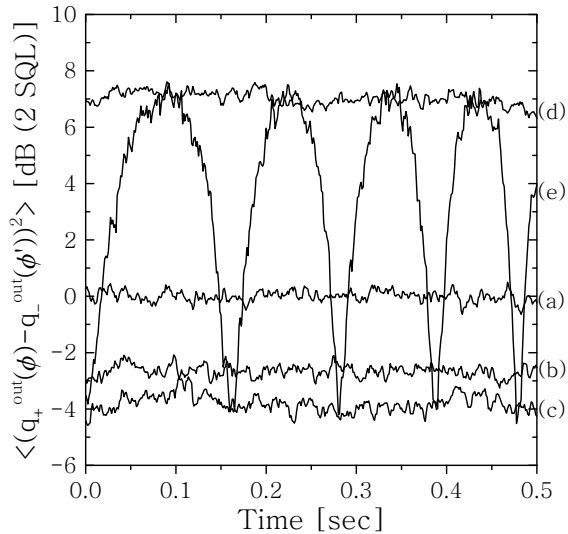}
\caption{\label{ndcorr}Quantum fluctuations of the signal/idler quadrature difference taken at $\psi=\psi_0$ and with various local oscillator phase settings. (a) Standard quantum limit for \textit{two} independent coherent states. Everything is normalized to this level (2 SQL). (b) Standard quantum level for a \textit{single} coherent state. (c) The correlated observable $(Q_+^{out})_{\psi_0}\simeq q_+(\phi_+)-q_-(\phi_-)$. (d) Anti correlated observable $(Q_-^{out})_{\psi_0}\simeq q_+(\phi_+)+q_-(\phi_-)=q_+(\phi_+ +\pi/2)-q_-(\phi_- +\pi/2)$. (e) $\phi$ kept constant while $\phi'$ was scanned (almost) linearly in time. The traces are taken at 1.15 MHz with a resolution bandwidth of 30 kHz.}
\end{figure}

With our OPO characterized in terms of its ability to produce quadrature
phase squeezed light we could now move on to investigate the quantum
correlations between the frequency non degenerate OPO cavity modes adjacent
to the degenerate mode. Since these modes only were shifted $\pm 370$ MHz
relative to the degenerate mode it was safe to assume equal output coupler
transmissions and internal OPO losses for both modes. However, since the
signal and idler fields experienced different optical paths, the detection
efficiencies were not equal for the two modes, meaning that $\sigma \neq 1$
in (\ref{genfluc}), whereas we had $\rho =1$. The limiting factors on the
inter beam correlations are listed in Table \ref{tab:epr}, from which we
derive the parameters $\xi =\sqrt{\xi _{+}\xi _{-}}=0.74\pm 0.01$ and $\sigma =\sqrt{\xi _{+}/\xi _{-}}=0.98\pm 0.01$. 
\begin{table}
\begin{ruledtabular}
\begin{tabular}{lcc}
Loss source & Signal & Idler \\ 
\hline
$\eta_{\pm}$ & 0.847 & 0.847\\ 
$\eta^{LO}_{\pm}$ & $(0.98)^2$ & $(0.975)^2$\\ 
$\eta^{QE}_{\pm}$ & 0.98 & 0.98\\
$\eta^{prop}_{\pm}$ & 0.91 & 0.96 \\
\hline
$\xi_{\pm}$ & $0.73\pm 0.02$ & $0.76\pm 0.02$\\
\end{tabular}
\end{ruledtabular}
\caption{The detection efficiencies for the ND-OPO signal (+) and idler (-)
modes.}
\label{tab:epr}
\end{table}
By using (\ref{optphase}) we find the optimum phase choice $\psi _{0}=-\pi
/4+0.0086$ radians which inserted in (\ref{genfluc}) gives a predicted
maximum degree of inter beam correlations of -4.9 dB. It should be mentioned that the correction to $\psi$ due to asymmetric losses was so small that we chose not to make any at all. This also applied to the correction of $\psi$ due to a finite degree of correlations. Hence within our uncertainties we used $\psi=\pi/4$ throughout the experiment. The measured correlations are shown in Fig. \ref{ndcorr}, from which we find the noise
reduction relative to two independent coherent state correlations of $-3.8\pm 0.1$ dB. Correcting again for electronic noise we get $-4.3\pm 0.3$
dB noise reduction due to the inter beam quantum correlations. Again this is
in reasonable agreement with our theoretical predictions. 
Fig. \ref{ndcorr} shows five traces: (a) The SQL obtained by blocking the
ND-OPO pump, hence only the vacuum input to the heterodyne detectors was
being probed yielding the quantum fluctuations from two uncorrelated vacuum
states. (b) The SQL from a single heterodyne detector. This level was found
by blocking the local oscillator of the other heterodyne detector as well as
the ND-OPO pump. Unblocking now all beams we obtained the trace (c).
As documented by the trace, phase fluctuations in our setup were relatively
small, so that the local oscillator phases could be maintained constant on a
timescale of hundreds of milliseconds in a way that maximum inter-beam
quantum correlations were observed. The same procedure was used to
obtain the level of maximum anti correlations (d). Next one of the local
oscillators phases was allowed to drift freely while the other phase
was scanned at a high rate. This resulted in the oscillatory trace (e)
showing the transition between observing the correlated  $\left\langle (Q_+^{out})^2 \right\rangle $ and the anti correlated $\left\langle (Q_-^{out})^2 \right\rangle$ defined in (\ref{crossqdef}).

According to the criterion for state separability developed in refs. \cite{simon00,duan00a} we must have
\begin{equation}
\begin{array}{l}
\left\langle (q_+(\phi_+)-q_-(\phi_-))^2 \right\rangle + \\
\left\langle (q_+(\phi_+ +\pi/2)+q_-(\phi_- +\pi/2))^2 \right\rangle<2
\end{array}
\label{separ}
\end{equation}
in order to be able to claim state inseparability for the signal and idler fields and hence entanglement. Inserting the measured values of interbeam correlations we get $0.37+0.37=0.74<2$. Hence our signal and idler fields are indeed in an entangled quantum state. Since the right hand side of (\ref{separ}) sets the limit for separability we can quantify our degree of entanglement as -4.3 dB relative to this limit. To our knowledge this is strongest entanglement demonstrated up to now using a narrowband tunable lightsource. We are of course aware that no universal measure of the strength of entanglement has been established yet, hence our quantification above only applies for the Gaussian continuous variable case.

\begin{figure}
\includegraphics{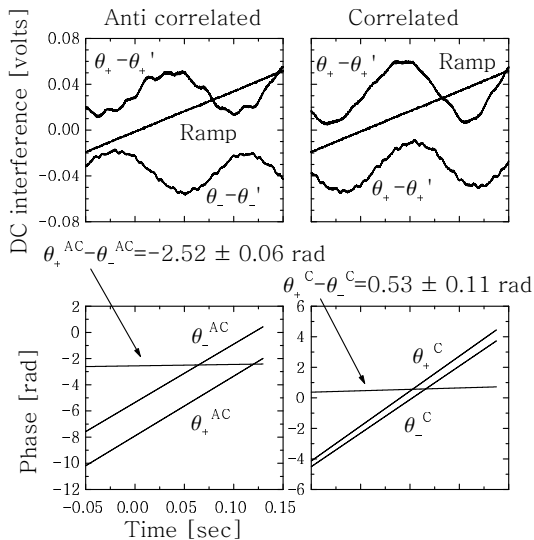}
\caption{\label{fig:dcphase}The DC interference fringes revealing the relationship between the local oscillator phases and the ND-OPO pump field. The left column corresponds to the situation when the local oscillator phases were locked to produce maximum anti correlations in $(Q_-^{out})_{\psi_0}$, while the right column are the results when maximum correlations were observed. The upper row are the raw data and the lower row are the local oscillator phases and their differences inferred from sinusoidal fits to the raw data.}
\end{figure}

In order to get a measure of the local oscillator phases we now injected a weak coherent field into the ND-OPO through a high reflector. The purpose of this exercise was to show that the sum of the local oscillator phases must be shifted by $\pi$ radians in order to switch from observing correlations to observing anti correlations as illustrated on Fig. \ref{fig:phasor}.  The frequency of the injected field was $\omega_-$, hence a coherent field with the frequency $\omega_+$ was produced in the ND-OPO via difference frequency generation. Given that the injected field had the phase $-\theta'$ relative to half the pump phase $\chi/2$ we know that the high frequency field was produced with the phase $+\theta'$ relative to $\chi/2$. As a result we could by observing the DC interference fringe between these two fields and the respective local oscillators learn about the local oscillator phase settings relative to the ND-OPO pump phase. An example of the interference fringes is shown in Fig. \ref{fig:dcphase}. In order not to disturb the quantum noise measurements taking place in the heterodyne detectors the coherent amplitudes of the injected field had to be small. Hence only about 100 nW hit the detectors and consequently the DC interference fringe visibility was only about 0.01 with a rather poor signal to noise ratio as a result. Again we left one local oscillator phase freely drifting but this time the other local oscillator phase was actively stabilized to produce either maximum or minimum inter beam quantum correlations. This was done using a standard dither and lock approach with an error signal derived from the spectrum analyzer output. Now the phase of the injected field was scanned linearly in time and the resulting interferences with the local oscillators were recorded. Before commenting on the results displayed in Fig. \ref{fig:dcphase} we should note that when $\theta'$ was varied in the negative direction for the $\omega_-$ field, the change of $\theta'$ would be in the positive direction for the $\omega_+$ field due to phase conservation in the difference frequency generation process. Hence we would expect the phase \textit{difference} between the two interference fringes to be constant when the local oscillator phases were fixed to produce maximum or minimum quantum correlations between the two heterodyne detectors. If this phase difference was constant it would correspond to having the local oscillator phases adding up to a constant value as illustrated in Fig. \ref{fig:phasor}.
We found the phase differences $\theta_{\pm}^{C}=\theta_{\pm}-\theta_{\pm}'$ and $\theta_{\pm}^{AC}=\theta_{\pm}-\theta_{\pm}'$ for the two frequencies in the cases of being locked to maximum correlations and maximum anti correlations respectively. This was done by fitting the somewhat noisy interference fringes to a sinusoidal function and the resulting phase variations as well as their differences are shown on the graphs of Fig. \ref{fig:dcphase}. We found $\theta_+^{C}-\theta_-^{C}=0.53\pm 0.11$ rad and $\theta_+^{AC}-\theta_-^{AC}=-2.52\pm 0.06$ rad, and we note that $(\theta_+^{C}-\theta_-^{C})-(\theta_+^{AC}-\theta_-^{AC})=3.05\pm 0.13$ rad. Ideally the last result should be $\pi$ since the sum of the local oscillator phases should be shifted by this amount in order observe the change from measuring a quiet to a noisy quadrature combination. As a final remark here we note that our setup was one big Mach-Zender interferometer from which the systematic phaseshift of 0.53 radians seen in $\theta_+^{C}-\theta_-^{C}$ was originating. The procedure described here for establishing the local oscillator phases was used in the following experiment, where the EPR paradox was revived for a short while.

\subsection{The EPR paradox revisited}

It is well known that the quantum correlated field quadratures of the ND-OPO
output can be used to visualize the EPR paradox in its original form put
forward by Einstein, Podolsky and Rosen in 1935 \cite{einstein35}. This has
been first demonstrated using a polarization non degenerate OPO \cite{ou92a,ou92b}. It is known that the non-negativity of the Wigner function
describing the ND-OPO output prohibits us from violating local hidden
variable theories when measuring quadrature phase amplitudes \cite{bell87}.
However an experimental realization of the EPR paradox is still interesting
especially seen in the light of recent proposals suggesting that Bell
inequalities generalized to continuous variables can be violated by
measuring parity like observables \cite{chen2001,banaszek98,banaszek99}.
Moreover when considering quantum information transfer with continuous
variables using the ND-OPO, like quantum teleportation \cite{furusawa98} one
must have a fidelity larger than 2/3 for outperforming classical systems
consisting of linear optics and coherent states in terms of reproducing the input state \cite{grosshans01}. This can only
be achieved when more than 50\% noise reduction is present in the
correlation between the signal and idler fields, which is just the condition
required for realizing the EPR paradox \cite{reid89}. We note that the 50\% limit only applies when one uses $\psi=\pi/4$. For other values of $\psi$ it turns out that the EPR paradox can be realized for a smaller degree of correlations. However demonstration of
the EPR paradox with our choice of $\psi$ in our ND-OPO is equivalent to demonstration of its
usefulness as a resource for continuous variable quantum information transfer like
cryptography \cite{ralph99,ralph00}, teleportation \cite{vaidman94}, storage \cite{polzik99} etc.
\begin{figure}
\includegraphics{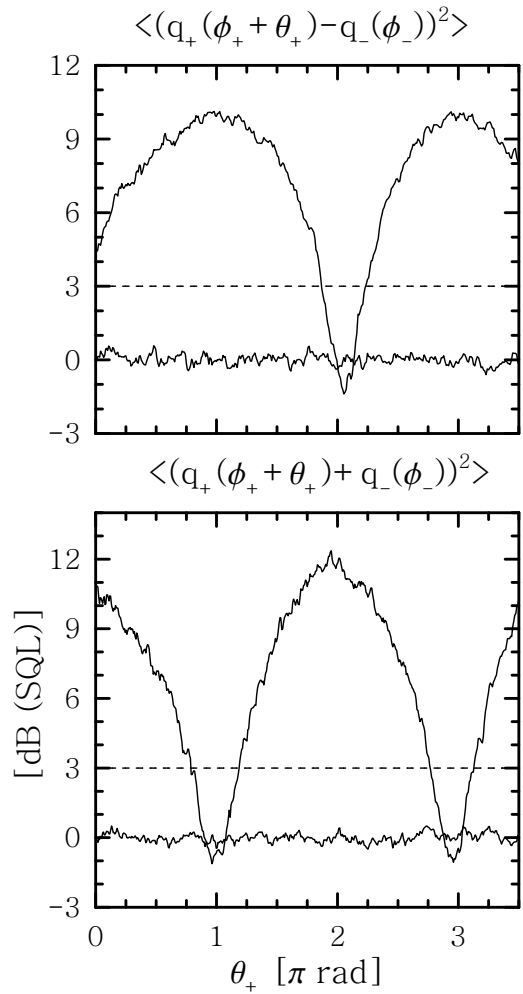}
\caption{\label{fig:epr}The noise densities in the quadrature difference (upper trace) and sum (lower trace) obtained while the relative local oscillator phase was scanned. Both traces are normalized to the SQL of a \textit{single} coherent state. Here the 3 dB lines indicate the SQL of two coherent states.}
\end{figure}

We refer to refs. \cite{peres95} and \cite{ou92a,ou92b} for the detailed
discussion of the EPR paradox. In brief we arrive at the paradox by
inferring the idler quadratures $q_{-}(\phi _{-})$ and $q_{-}(\phi _{-}+\pi
/2)$ from measuring the correlated signal variables $q_{+}(\phi _{+})$ and $q_{+}(\phi _{+}+\pi /2)$. One can choose at will to measure either of the
two variables, hence according to EPR they both should be "elements of
objective reality". However according to quantum mechanics the non commuting
observables $q_{-}(\phi _{-})$ and $q_{-}(\phi _{-}+\pi /2)$ cannot have
objective values simultaneously, hence the paradox, which is demonstrated if
the product of the inferred uncertainties is $\left\langle (q_-(\phi_-))^2 \right\rangle _{inf} \left\langle (q_-(\phi_- +\pi/2))^2 \right\rangle _{inf} < 1$.

Turning now to the experiment we sent the photocurrents from the two
heterodyne detectors into a hybrid junction producing the sum and the
difference of the photocurrents in its two outputs. These outputs were fed
into two RF spectrum analyzers in order to establish the noise density
around 1.15 MHz. During this experiment one local oscillator phase was kept
constant while the other was scanned linearly. Like in the previous section
we also here injected a weak signal into the OPO in order to keep track of
the local oscillator phases. Depending on the phase of scanned local
oscillator the outputs of the hybrid junction yielded the observables $q_{+}(\phi _{+}+\theta _{+})\pm q_{-}(\phi _{-})$. That is, with $\theta
_{+}=0$ we measured the non commuting observables $q_{+}(\phi
_{+})-q_{-}(\phi _{-})$ in the (-) junction and $q_{+}(\phi _{+})+q_{-}(\phi
_{-})=q_{+}(\phi _{+}+\pi /2)-q_{-}(\phi _{-}+\pi /2)$ in the (+) junction
(see Fig. \ref{fig:setup}). When $\theta _{+}=\pi $ we measured $-(q_{+}(\phi
_{+})+q_{-}(\phi _{-}))$ in the (-) junction and $-(q_{+}(\phi
_{+})-q_{-}(\phi _{-}))=-(q_{+}(\phi _{+}+\pi /2)+q_{-}(\phi _{-}+\pi /2))$
in the (+) junction. From (\ref{symmopo}) we see that with $\theta _{+}=0$
we measured a quiet observable in the (-) junction and hence gained information
about $q_{-}(\phi _{-})$ while with $\theta _{+}=\pi $ we measured a quiet
observable in the (+) junction and then learned about the conjugate idler
observable $q_{-}(\phi _{-}+\pi /2)$. The result of a scan is shown in Fig. \ref{fig:epr}. Here the fluctuations of the two measurements are shown
normalized to the vacuum noise level of a \textit{single} detector. After
averaging over 12 pairs of traces like the ones displayed in Fig \ref{fig:epr} we found the uncertainties  $\left\langle (q_+(\phi_+)-q_-(\phi_-))^2 \right\rangle=-1.0\pm 0.1$ dB and $\left\langle (q_+(\phi_+ +\pi/2)+q_-(\phi_- +\pi/2))^2 \right\rangle=-0.9\pm 0.1$ dB both relative to
the single beam SQL. These uncertainties resulted in an error of the same
size in our estimation of the idler quadratures. Hence we could readily
calculate the product of inferred variances in this experiment to be $\left\langle (q_-(\phi_-))^2 \right\rangle_{inf} \left\langle (q_-(\phi_- +\pi/2))^2 \right\rangle_{inf}=0.65\pm 0.02<1$ which was a clear demonstration of the paradox.

\section{Summary}

We have demonstrated how the frequency non degenerate optical parametric
oscillator can be used to generate spatially separated entangled optical
fields. In this paper we have used the OPO cavity resonances adjacent to the
frequency degenerate mode, resulting in entangled fields separated by 740
MHz corresponding to two OPO cavity free spectrum ranges. In principle one
could select any pair of modes placed symmetrically around the degenerate
mode, as long as the $\chi ^{(2)}$ phase matching condition is fulfilled for
this pair of modes. The entanglement for our source is present over a bandwidth of approximately 14 MHz FWHM depending on ND-OPO pump parameter $\varepsilon$. This results in a high spectral brightness making the source well suited for addressing atomic transitions.

In particular our source can be used for quantum memory readout via interspecies teleportation of a collective atomic spin state onto the polarization state of beam of light \cite{kuzmich00}. For this purpose two modes of light displaying entanglement in the Stokes parameters would be needed and only minor modifications to our source would be needed in order produce such states of light.

We found a degree of squeezing of -4.0 dB of the frequency degenerate
mode and we found inter beam correlations of the nondegenerate modes of -3.8
dB. Both results could be improved by approximately 0.5 dB if the thermal
noise of our detectors could be suppressed. In this case the agreement with
our theoretical predictions was reasonable.

To gain information about the OPO pump phase we injected a weak
coherent field with frequency $\omega _{-}$ and thereby produced a $\omega
_{+}$ field via difference frequency generation. These fields interfered
with the local oscillators yielding us information about the local
oscillator phases relative to the OPO pump field. Here we found that the sum
of local oscillator phases changed by $3.05\pm 0.13$ rad. when going from
observing maximum inter beam correlation to observing maximum anti
correlations. This was consistent with the theoretical value of $\pi $.

Finally we used the entangled fields of our ND-OPO to visualize the
Einstein-Podolsky-Rosen paradox in its original form with continuous quantum
variables. Here we found a product of inferred variances of $\left\langle (q_-(\phi_-))^2 \right\rangle_{inf} \left\langle (q_-(\phi_- +\pi/2))^2 \right\rangle_{inf}=0.65\pm 0.02$ which
shows the paradox since this is significantly smaller than the Heisenberg limit of unity.

\begin{acknowledgments}
This work is supported by the Danish National Research Foundation and by the EU
QIPC network via the QUICOV project. We wish to thank O. Arcizet for
interesting discussions and his help at the initial stage of the experiment.
\end{acknowledgments}

\bibliographystyle{apsrev}
\bibliography{epr11}

\end{document}